\documentclass[conference]{IEEEtran}
\IEEEoverridecommandlockouts

\usepackage{amsmath,amssymb,amsfonts}
\usepackage{algorithmic}
\usepackage{graphicx}
\usepackage{textcomp}

\usepackage{epsf}
\usepackage{latexsym}
\usepackage{subfigure}
\usepackage{multirow}
\usepackage{multicol}
\usepackage[table]{xcolor}
\usepackage{capt-of}

\usepackage[nospace,noadjust]{cite}

\usepackage{algorithm}

\begin{document}

\title{Mitigation of Human Exposure to RF Fields in Downlink of Millimeter-Wave Cellular Systems}

\author{\IEEEauthorblockN{Imtiaz Nasim and Seungmo Kim}
\IEEEauthorblockA{The Electrical and Computer Engineering Department\\
Georgia Southern University\\
Statesboro, GA 30460, USA \\
\{in00206, seungmokim\}@georgiasouthern.edu}
}

\maketitle

\begin{abstract}
Out of the very few studies that paid proper attention to the harmful health impacts in millimeter-wave (mmW) communications, most of them are concerned about uplink cases due to closer contact with the human body. Our recent study revealed that even the human exposure to radio frequency (RF) fields in downlink mmW technology is not very minimum to be ignored. There were a few RF exposure mitigation techniques for uplinks, but the downlink scenario is hardly paid any attention. However, this paper proposes a downlink protocol for mmW cellular communications that achieves the maximum data rate while keeping the impacts on human health minimized. Our results show that the proposed technique lowers both power density (PD) and specific absorption rate (SAR) compared to the typical protocol, with only slight sacrifice in data rates.
\end{abstract}


\section{Introduction}\label{sec_intro}
The Fifth Generation Wireless Systems (5G) have gained a huge research interest as a promising solution for the existing bandwidth shortage and lower data rate problems for future communications technology. But due to the probable implementation of \textit{narrower beams} with highly directive antenna arrays \cite{agiwal16}-\cite{shakib16} and \textit{larger number of transmitters} have the potential to increase the  concern of higher RF exposure to human users at mmW communications. Moreover, more base stations (BSs)/access points (APs) are likely to be deployed closer to each other in 5G \cite{sung17}-\cite{tr38900}, compared to the present communications architecture. This will increase the level of human exposure to RF radiations. The advancement in RF circuits is able to integrate larger number of miniaturized antennas which can produce higher antenna gains.

\subsection{Related Work}\label{sec_related}
 This paper is motivated from the fact that prior work is not enough to address such potential threats of RF exposure in cellular communications networks.

\subsubsection{Measurement of Human RF Exposure}
The Federal Communications Commission (FCC) \cite{fcc01} and International Commission on Non-Ionizing Radiation Protection (ICNIRP) \cite{icnirp98} set the maximum allowable limit for electromagnetic (EM) emission radiation that can be allowed to penetrate into the human body without causing much health concerns and protect the human users from the undesirable effects of wireless radiation. The aforementioned features of 5G--(i) smaller AP-UE distance and (ii) narrower beam--can threaten the human health by higher RF exposure even at downlinks \cite{icc}. This claims that the downlink exposure cannot be ignored and hence RF mitigation techniques are also required in the downlink for mmW technology to ensure a safe communications environment.

Possibilities of skin cancer due to RF emissions at higher frequency spectrum are reported \cite{gao12}. Heating due to EM exposure in mmW is absorbed within the first few millimeters (mm) within the human skin; for instance, heat is absorbed within 0.41 mm for 42.5 GHz \cite{alekseev01}. The mmW induced burns are more likely to be conventional burns as like as a person touching a hot object as reported in \cite{wu15}. The normal temperature for the skin outer surface is typically around 30 to 35$^{\circ}$C. The pain detection threshold temperature for human skin is approximately 43$^{\circ}$C as reported and any temperature over that limit can produce long-term injuries. 

One problem is that the literature on the impact of cellular communications on human health is not mature enough. The three major quantities used to measure the intensity and effects of RF exposure are SAR, PD, and the steady state or transient temperature \cite{em92}\cite{em05}. However, selection of an appropriate metric evaluating the human RF exposure still remains controversial. The FCC suggests PD as a metric measuring the human exposure to RF fields generated by devices operating at frequencies higher than 6 GHz \cite{fcc01}, whereas a recent study suggested that the PD standard is not sufficient to determine the health issues especially when devices are operating very close to human body in mmW \cite{rappaport15}. Therefore, this paper examines the human RF exposure by using both PD and SAR.

\subsubsection{Reduction of Human RF Exposure}
Very few prior studies paid attention to human RF exposure in communications systems \cite{wu15}\cite{rappaport15}-\cite{chahat12}. Propagation characteristics at different mmW bands and their thermal effects were investigated for discussion on health effects of RF exposure in mmW radiation \cite{rappaport15}. Emission reduction scheme and models for SAR exposure constraints are studied in recent work \cite{love16}\cite{sambo15}.

However, health impacts of mmW RF emissions in \textit{downlink} of a cellular communications system have not been studied thoroughly. Our recent work in \cite{icc} suggests even the downlink communications can also produce significant radiation level which is very much capable of producing potential threat to human health at mmW spectrum. Moreover, our study also urges the necessity of considering SAR for far-field downlink in mmW communications for determining the health impacts. As the impact of radiation in downlink cannot be ignored, there remains a strong necessity for the development of RF mitigation in mmW bands for the successful deployment of 5G communications. If the radiation level for both uplink and downlink can be maintained at a tolerable range following the restriction guidelines, only then the future communications model will go on to serve the user with its maximum efficiency and smart features with enhanced service quality.

\subsection{Contributions}
Three contributions of this paper can be highlighted and distinguished from the prior art.

Firstly, extending our prior work \cite{icc}, this paper analyzes the human RF exposure in the \textit{downlink} more thoroughly. All the prior work studied an uplink only, while paid almost no attention to suppression of RF fields generated by APs. In fact, the APs generate even stronger RF fields compared to the concurrent systems, due to (i) higher transmit power and (ii) larger antenna array size leading to higher concentration of RF energy. Moreover, one important feature of the future cellular networks is small cell networks. The consequences of this change will be two-fold: (i) APs/BSs will serve smaller geographic areas and thus are located closer to human users; (ii) larger numbers of APs/BSs will be deployed, which will lead to higher chances of human exposure to the RF fields generated by downlinks.

Secondly, this paper \textit{proposes a downlink protocol that suppresses the human RF exposure}. It elects the serving AP for a UE among the ones with the SAR below the FCC's guideline at the carrier frequency of 28 GHz \cite{fcc01}. That says, while the typical downlink connects a UE to the AP with the strongest received signal strength among all of the APs around, the proposed protocol selects one among the APs keeping the SAR at safe levels.

Thirdly, this paper uses the available PD regulations for mmW spectrum set by FCC to produce new SAR values for downlink far-field cases. As SAR guidelines are also required to determine the health impacts \cite{icc}, these new SAR values may act as a guide to set up the regulations for cellular communications in downlinks.

\begin{table}[t]
\centering
\scriptsize
\caption{Parameters for 5G}
\begin{tabular}{|c|c|}
\hline
\textbf{Parameter} & \textbf{Value}\\ \hline \hline
Carrier frequency & 28 GHz\\
System layout & RMa, UMa, UMi \cite{tr38900}\\
Inter-site distance (ISD) & {200 m}\\
Cell sectorization & {3 sectors/site}\\
Bandwidth & {850 MHz}\\
Max antenna gain & {5 dBi per element}\\
Transmit power & 21 dBm per element\\
AP's number of antennas ($\lambda/2$ array) & 8$\times$8\\
AP antenna height & 10 m\\
Duplexing & {Time-division duplexing (TDD)} \\
Transmission scheme & {Singler-user (SU)-MIMO} \\
UE noise figure & {7 dB}\\
Temperature & {290 K}\\ \hline
\end{tabular}
\label{table_parameters}
\end{table}

\begin{figure}[t]
\centering
\includegraphics[width = 3.5in, height= 3in]{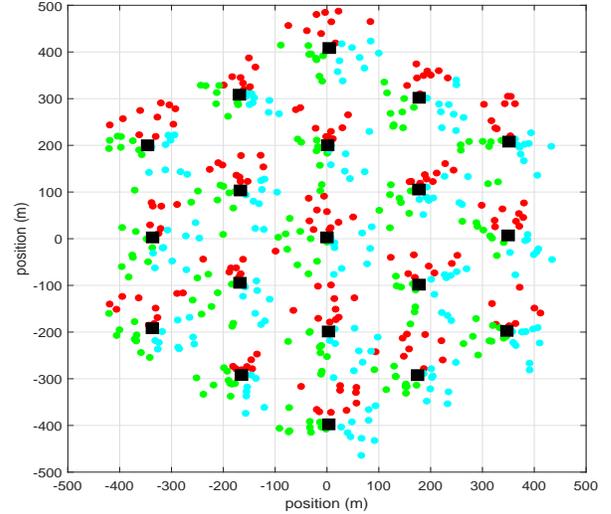}
\captionof{figure}{A snapshot of one ``drop'' of 5G topology (19 sites, 3 sectors per site, and 10 UEs per sector)}
\label{fig_topology}
\end{figure}

\section{System Model}\label{sec_system_model}
This section describes the system setting for a cellular communications network that forms the basis for analysis of human RF exposure. Considering the frequency spectrum of 28 GHz as a potential candidate for 5G, we use a corresponding technical report \cite{tr38900} that was released by the 3GPP. Also, this paper compares the human RF exposure level in a 5G system between the proposed protocol that selects an AP for a UE keeping the SAR below the FCC guideline of 10 W/kg, and the typical protocol which connects a UE to the AP with the strongest received signal. The parameters for a 5G network are summarized in Table \ref{table_parameters}.

\subsection{Path Loss}
Our model for a 5G network is illustrated in Fig. \ref{fig_topology}. It consists of 19 sites each having 3 sectors. The inter-site distance (ISD) is 200 meters (m) and each sector is assumed to have 10 active UEs. Also, as identified in Table \ref{table_parameters}, for the terrestrial propagation between an AP and a UE, the following three path loss models are assumed: Rural Macro (RMa), Urban Macro (UMa), and Urban Micro (UMi) \cite{tr38900}.

Though we chose the carrier frequency of 28 GHz to design our model, the analysis framework can be extended and the performance can be demonstrated for any other standards of cellular networks, following our methodology. The model has random UE location and random line-of-sight (LoS) for each and every UE to make it more realistic with the real time cases. It should be noted that the present technology is composed of larger cells wherein a single BS can provide coverage up to several kilometers (km), which is in contrast to a 5G network operating at higher frequencies (i.e., 28 GHz), adopting relatively smaller cells. As such, in 5G, the same area is covered by a larger number of APs in denser deployment in order to provide, mainly due to faster attenuation of EM waves.

\subsubsection{Antenna Beam Pattern}
For a 5G AP, the attenuation patterns of an antenna element on the elevation and azimuth plane are given by \cite{tr38900}
\begin{align}
\label{eq_antenna_5g_a}
A_{a}\left(\phi\right) &= \min\left\{12\left(\frac{\phi}{\phi_{3db}}\right)^2, A_m\right\} \rm{~[dB]}\\
\label{eq_antenna_5g_e}
A_{e}\left(\theta\right) &= \min\left\{12\left(\frac{\theta-90^{\circ}}{\theta_{3db}}\right)^2, A_m\right\} \rm{~[dB]}
\end{align}
where $\phi$ and $\theta$ are angles of a beam on the azimuth and elevation plane, respectively; $\left(\cdot\right)_{3db}$ denotes an angle at which a 3-dB loss occurs. Then the antenna element pattern that is combined in the two planes is given by
\begin{align}\label{eq_antenna}
A\left(\theta,\phi\right) = \min \left(A_{a}\left(\phi\right)+A_{e}\left(\theta\right), A_m\right)\rm{~[dB]}
\end{align}
where $A_m$ is a maximum attenuation (front-to-back ratio). It is defined $A_m = 30$ dB in \cite{tr38900}, but it can be higher in practice. Finally, an antenna gain that is formulated as
\begin{align}\label{eq_geometry_G_final}
G\left(\phi,\theta\right)=G_{max} - A\left(\phi,\theta\right) \rm{~[dB]}
\end{align}
where $G_{max}$ is a maximum antenna gain.

\begin{figure*}[t]
\centering
\includegraphics[width = \linewidth]{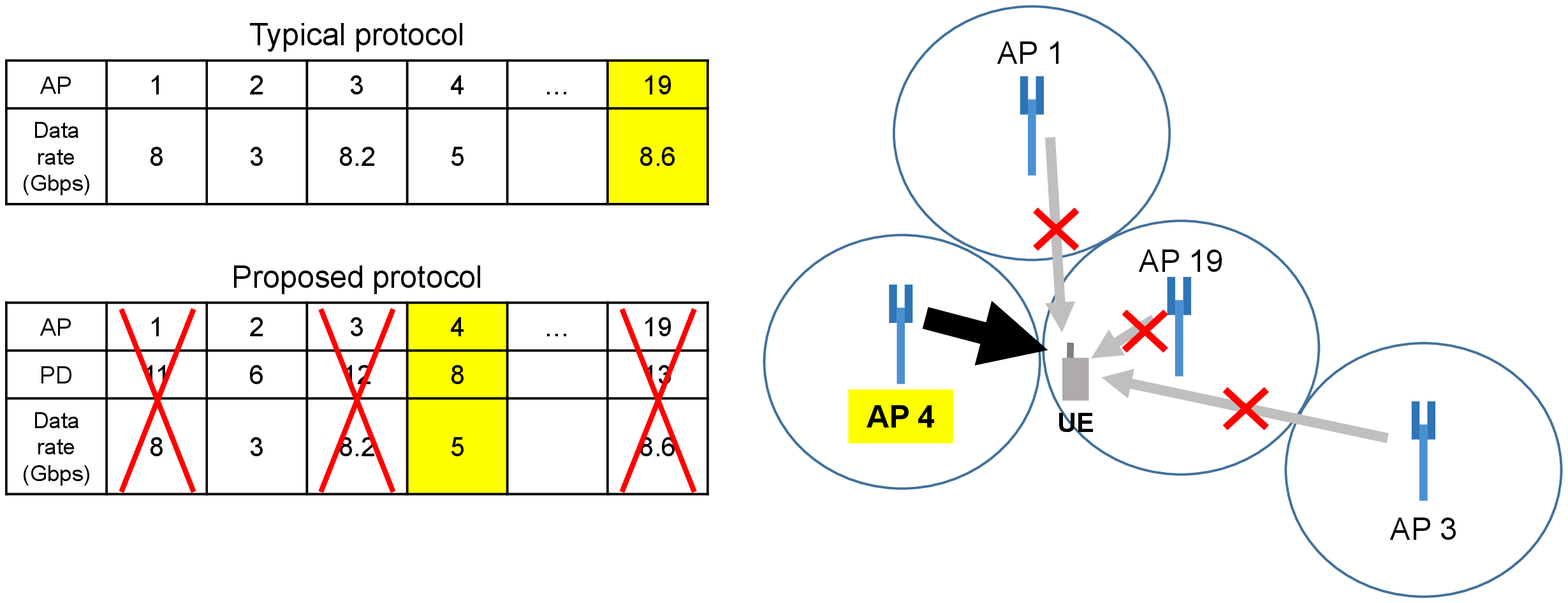}
\caption{An example usage of the proposed protocol}
\label{fig_protocol}
\end{figure*}

\section{Performance Analysis}
In this section, we present the analysis of our work to present a model that will reduce RF emissions at the user end in mmW systems. As explained in Section \ref{sec_system_model}, the smaller cell size in 5G leads to smaller ISD, which produces closer location of the human users with the serving APs within the cell for downlink operations.

\subsection{Data Rate}
The performance of a 5G downlink is represented by a data rate, which is given by
\begin{equation}\label{eq_rate}
R = B \log (1+ \gamma)
\end{equation}
where $R$ and $B$ denote the data rate and the bandwidth, respectively. A downlink signal-to-noise ratio (SNR) received at a UE is denoted by $\gamma$ which is calculated considering random locations of the UEs in a sector that is formed by an AP in a 5G system, as illustrated in Fig. \ref{fig_topology}. It is noteworthy that an accurate three-dimensional distance is considered, taking into account the exact heights of an AP and UE \cite{tr38900} as provided in Table \ref{table_parameters}.

\subsection{Human RF Exposure}
SAR and PD are considered as the most commonly used evaluation parameters so far to determine the deleterious impacts of RF emissions from communications network \cite{icc}. We show our analysis for both PD and SAR as there remains a controversy about which evaluation parameter is more accurate to be considered. To the best of our knowledge, there is no limit set for SAR for frequency spectrum over more than 6 GHz in downlink. We thus use the available PD restrictions according to the FCC guidelines for downlink cases with the carrier frequency higher than 6 GHz. The calculated PD values for each UE under the proposed protocol is then used to calculate the corresponding SAR for each UE. Thus, this paper infers a guideline on the SAR using the PD regulations, and applies it in downlinks for mmW communications.

A PD from a transmitting antenna for far-field \cite{wu15} can be expressed as,
\begin{equation}\label{eq_pd}
PD = \frac{|E_i|^2}{\eta} = \frac{\eta}{|H_i|^2}
\end{equation}
where \textit{$E_i$} and \textit{$H_i$}(A/m) denotes the root mean square (rms) values of the electric and magnetic field strengths in voltage per meter (V/m) and ampere per meter (A/m), respectively, incident on the exposed tissue surface and $\eta$ represents the wave impedance in ohm ($\Omega$) unit. PD is a measurement of the power dissipated per area of the exposed body tissue, whose unit is $\text{W/m}^2$. 

Another RF exposure evaluating parameter that is used most commonly is SAR. It is a quantitative measure representing the power dissipated per body mass of the exposed tissue and the SI unit of SAR is W/kg, which is calculated by
\begin{equation}\label{eq_sar}
SAR = \frac{P_{diss}}{m} = \frac{\sigma|E|^2}{\rho}
\end{equation}
where $P_{diss}$ represents dissipated power in the exposed tissue in the unit of Watts (W), m represents the mass of the exposed tissue in the unit of kg, $\sigma$ is the conductivity in siemens per meter (S/m), $\rho$  is the tissue mass density ($\text{kg/m}^2$) and \textit{E} is the rms value of electric field strength which is given in the unit of V/m. The value of SAR at the surface of an exposed tissue is different from one measured in the deeper tissue. Also, the value of SAR for a particular tissue in human body is different according to different body location--i.e., hands, head, etc.

This paper focuses on the downlink behaviors of the mmW system when performing analysis for the RF mitigation protocol. Incident PD for far-field downlink communications is expressed as 
\begin{equation}\label{eq_pd}
S_i = (P_tG_t)/(4\pi d^2)
\end{equation}
where $P_t$ represents the transmit power, $G_t$ is the transmitter antenna gain and $d$ denotes the AP-UE distance (m).

Now, we can rewrite SAR given in (\ref{eq_sar}) in terms of $d$ for calculation in a cellular communications system, which is also a function of $\phi$ \cite{love16} as,
\begin{equation}\label{eq_sar_d}
SAR (d) = SAR (\phi) = 2Si (\phi) T (\phi)m(\phi) / (\delta\rho)
\end{equation}
where \textit{T} is the power transmission coefficient \cite{love16} and $\delta$ is the skin penetration depth (m) measured at 28 GHz \cite{wu15}.The function $m(\phi$) is dependent on the tissue properties of dielectric constant ($\epsilon$*).

\section{Proposed Protocol}\label{sec_proposed_protocol}
For the RF mitigation in mmW downlinks, we propose a protocol that selects the serving AP for a UE, guaranteeing the PD under the FCC guideline. The guideline suggests the allowable limit for PD for frequency spectrum greater than 6 GHz as 10 $\text{W/m}^2$ \cite{fcc01}.

\begin{figure}[t]
\centering\includegraphics[width = \linewidth]{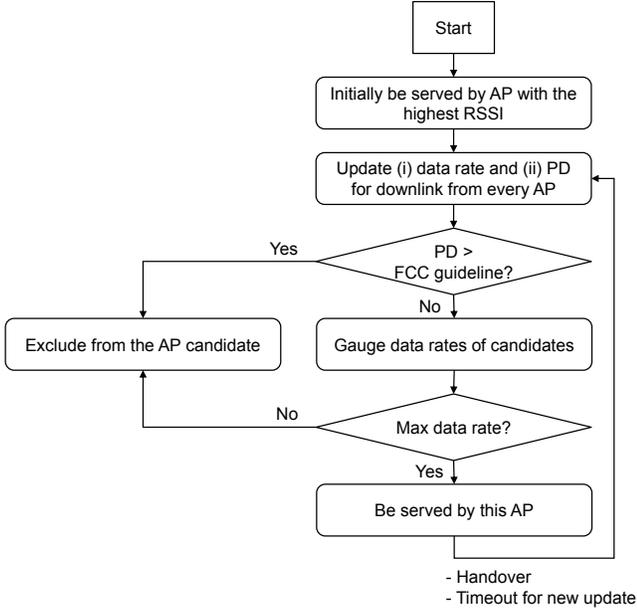}
\caption{Flowchart for the proposed protocol}
\label{fig_protocol_flowchart}
\end{figure}

Fig. \ref{fig_protocol_flowchart} provides a flowchart for the proposed protocol. Each UE is initially served by the AP with the RSSI, as in typical downlink protocols. However, the proposed protocol lets the UE update the PD as well, when it updates the information of the surrounding APs for purpose of possible handovers. This update is accomplished via a downlink pilot message. This PD level caused by each AP is used to examine whether it violates the FCC guideline, which is stored in the read-only memory (ROM) of each UE device according to the carrier frequencies at which it is supposed to operate. Among the APs with PDs under the FCC guideline, one providing the maximum downlink data rate is selected as the serving AP. This AP serves the UE until it (i) needs to be handed over to another cell or (ii) is served until a timeout. This timeout is set to periodically measure the PD again and select a new AP if the current serving AP comes to violate the guideline as the UE moves.

One key benefit of using PD as the metric to represent the human RF exposure level is that it can (i) directly  lead to a SAR level according to (\ref{eq_sar}), and (ii) be meausred at an AP. If it were to be measured at a UE, a separate control channel is needed to feed the SAR measured at a human user back to the serving AP. However, exploiting the fact that a SAR is directly computed from a PD, we propose that an AP measures its PD and periodically updates via downlink by piggybacking on other downlink control channels. This leads to a smaller number of feedback overhead between a UE and its sering AP, which in turn results in an efficient cellular networking.

Fig. \ref{fig_protocol} shows the proposed protocol that aims to achieve the maximum data rate in a downlink while keeping the PD under the FCC's guideling \cite{fcc01}. The tables compare the proposed protocol to the typical donwlink protocol. In the proposed protocol, APs 1, 3, and 19 are excluded due to the PD that they cause. It is compared to a typical downlink protocol in which the serving AP for a UE is selected solely accoding to the achievable downlink rate.

\section{Numerical Results}\label{sec_results}
In this section, we analyze the results of the performance of our proposed protocol (where the serving AP is selected among the APs with SARs below the FCC guideline) and compare it to that of the typical protocol (where the serving AP is selected among all).

The comparison is made in terms of data rates, which implies the performance in terms of service quality. Then the impacts on the human health is compared in terms of PD and SAR between the proposed and typical protocols.

\subsection{Data Rate}
\begin{figure}[t]
\centering\includegraphics[width = \linewidth]{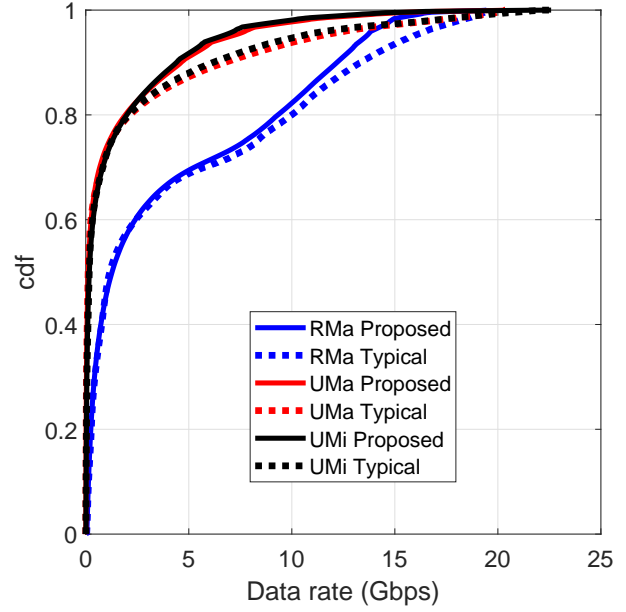}
\caption{Data rate comparison between the proposed protocol and typical protocol in terms of cdf}
\label{fig_rate}
\end{figure}

Fig. \ref{fig_rate} shows the comparison of data rates that can be achieved in a 5G mmW communications system between the typical protocol and the proposed RF mitigation protocol. It is shown from the figure that the proposed protocol sacrifices data rates as it prioritizes the SAR (to the data rate) in selection of the serving AP for a UE, as illustrated in Fig. \ref{fig_protocol}. In contrast, in the typical protocol, a UE selects the serving AP with the maximum received power, which leads to the maximum data rate. This AP may provide a PD value which is higher than the regulations and is capable to cause health concerns.

However, it should be noted from Fig. \ref{fig_rate} that, both protocols have a data rate which is in several multi-gigabyte-per-second (Gbps) range and thus falls in a desired level for the 5G. It indicates that our proposed protocol is still able to serve a downlink at a reasonable data rate, in spite of several Gbps of degradation. In general, 5G systems will be expected to provide a high degree of coverage and reliability even in the most severe propagation environments. In \cite{rappaport_jsac14}, an SINR both for uplink and downlink at mmW frequencies should be kept above -10 dB. This is interpreted to 0.1169 Gbps, with $B = $ 850 MHz as given in Table \ref{table_parameters} with substitution into (\ref{eq_rate}), which could be observed at the proposed protocol in Fig. \ref{fig_rate}. This indicates that the 5G systems are expected to remain fully operational with the proposed protocol adopted, even at the lowest-case downlink rate. Thus, we conclude that despite of some degradation in both downlink and uplink due to incumbent interference mitigation, the performance of a 5G system will remain acceptable.

\subsection{Human RF Exposure}
Now we show the performance of our proposed protocol in terms of RF exposure to human users from communications network in mmW downlink scenarios. Even after considering such shallow penetration depth due to high frequencies, downlink RF emissions can cause significantly higher level of PD and SAR if no mitigation technique is adopted. In this section, we compare our proposed protocol to the typical protocol and  urge the necessity of a RF mitigation model for mmW downlink communications. As there is no SAR guideline for RF exposure in far-field downlink cases according to the FCC guideline, we use the available PD guidelines to demonstrate the performance of our proposed model first. Then with the achieved PD values following the guideline, we achieve the corresponding SAR values which may  infer to set the new guidelines for SAR in mmW far-field downlink.

\begin{figure}[t]
\centering\includegraphics[width = \linewidth]{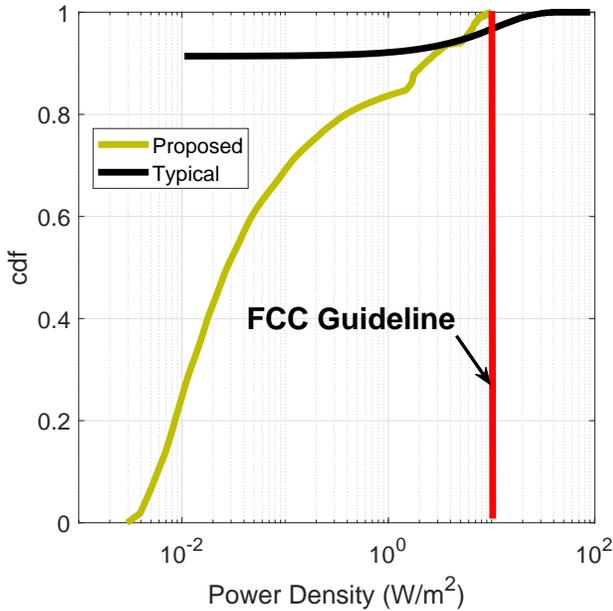}
\caption{Comparison of PD between the proposed protocol and typical protocol in terms of cdf}
\label{fig_pd}
\end{figure}

Fig. \ref{fig_pd} shows the PD comparison between our proposed protocol and the typical protocol. It can be seen that without any RF mitigation scheme adopted, the power density is very high for the existing UEs under the future mmW technology. In fact, more than 80 percent of the UEs are exposed to higher PDs than the guideline which is 10 $\text{W/m}^2$ for general public for frequecny spectrum ranging from 1.5-100 GHz according to FCC guideline \cite{wu15}. But adopting our proposed protocol, the probability of an UE being exposed to higher  PD than guideline falls to a significant level and stays within the limit. 

\begin{figure}[t]
\centering\includegraphics[width = \linewidth]{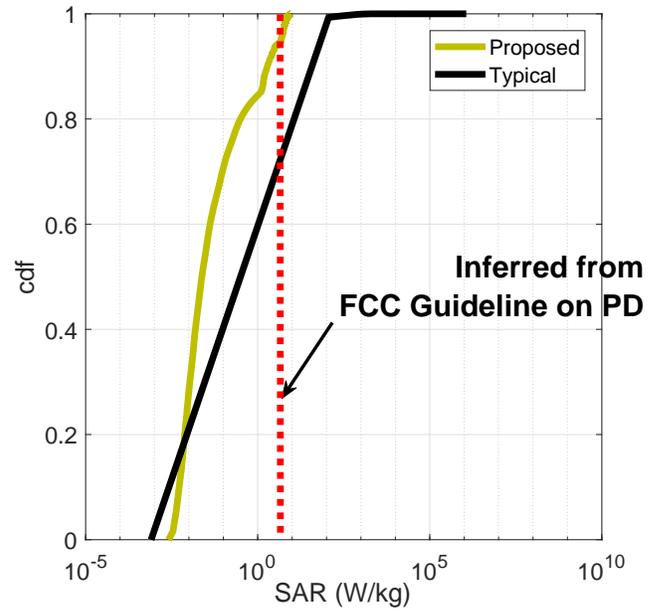}
\caption{Comparison of SAR between the proposed protocol and typical protocol in terms of cdf}
\label{fig_sar}
\end{figure}

Fig. \ref{fig_sar} compares the proposed scheme to the typical one in terms of SAR. It shows that the proposed protocol reduces not the SAR only but the variation over an entire cell. In other words, in the proposed protocol a 5G cell provides downlinks with SARs within 0.005 to 1, while the typical protocol yields a wider range spanning 0.001 to 100. Also, it is noteworthy that there is no guideline for SAR so far for the carrier frequency of 28 GHz, but our previous work \cite{icc} found the necessity of considering SAR even in downlink communications at such a high carrier frequency. The result observed from Fig. \ref{fig_sar} can support setting up future regulations/guidelines for mmW downlink communications in terms of SAR.

\section{Conclusions}\label{sec_conclusions}
Distinguished from the prior studies that focused on RF mitigation for uplinks only, this paper has highlighted the significance of RF radiation in downlinks and has proposed a novel downlink protocol that reduces RF exposure in a cellular network operating at a mmW frequency. This paper showed that the human RF exposure is likely to be increased in a 5G network due to adoption of larger phased array antennas. Our results showed our proposed protocol can reduce the human RF exposure level at the cost of downlink performance degradation. While depending on the path loss model, the proposed protocol still provides more than 90\% of the UEs in a cell with downlink data rates higher than 5 Gbps.


\end{document}